\documentclass[prl,twocolumn,aps,showpacs]{revtex4}
\usepackage{amssymb,color,graphicx,bm}

\definecolor{red}{rgb}{0.8,0,0}
\definecolor{RED}{rgb}{0.8,0,0}
\definecolor{violet}{rgb}{0.4,0,0.4}
\definecolor{green}{rgb}{0,0.5,0.0}
\definecolor{GREEN}{rgb}{0,0.5,0.0}
\definecolor{navy}{rgb}{0.0,0.0,0.6}
\definecolor{orange}{rgb}{0.8,0.2,0.0}
\definecolor{blue}{rgb}{0.3,0.0,0.8}

\begin{document}

\title{Comment on ``Strong constraints on magnetized white dwarfs surpassing the Chandrasekhar 
mass limit''}

\author{Upasana Das, Banibrata Mukhopadhyay\\
Department of Physics, Indian Institute of Science, 
Bangalore 560012, India\\ upasana@physics.iisc.ernet.in , bm@physics.iisc.ernet.in\\
}

\begin{abstract}
We show that the upper bound for the central magnetic field of a 
super-Chandrasekhar white dwarf calculated by Nityananda and Konar 
[Phys. Rev. D 89, 103017 (2014)] 
is completely erroneous. This in turn strengthens the argument in favor of the 
stability of the recently proposed magnetized 
super-Chandrasekhar white dwarfs. We also point out several other numerical errors 
in their work. Overall we conclude, based on our calculations, 
that the arguments put forth by Nityananda and 
Konar are fallacious and misleading.
\end{abstract}


\maketitle

Super-Chandrasekhar white dwarfs are currently in the limelight. This is, on one hand, due to the 
discovery of several peculiar, overluminous type Ia supernovae, 
e.g. SN~2006gz, SN~2007if, SN~2009dc, SN~2003fg \cite{nature,scalzo,hicken,yam,silverman,
taub}, which seem to invoke the explosion of super-Chandrasekhar white dwarfs having mass $2.1-2.8M_\odot$. 
This is all the more so, since Mukhopadhyay and his collaborators \cite{kundu,prd12,ijmpd12,prl13,apjl13,grf13,mpla14,jcap14} 
initiated to explain such overluminous type Ia supernovae by proposing the existence of highly magnetized
super-Chandrasekhar white dwarfs, which was further criticized by some other authors \cite{dong,chamel,nityakon,nk2}. 

The doubts raised by some of the authors \cite{dong,chamel} are regarding the stability of these super-Chandrasekhar 
white dwarfs, which, however, we have addressed and answered in detail in our latest works \cite{mpla14,jcap14}. 
Some authors, on the other hand, supported our work in their computations \cite{cheon,herrera,smoller}. 
In fact, Federbush et al. \cite{smoller}, by an extensive mathematical analysis, have shown 
the existence of stable magnetic star solutions, which include our super-Chandrasekhar white dwarfs
\cite{prl13}. 

Nityananda and Konar \cite{nityakon,nk2} have raised exactly the same question, as the earlier 
critics \cite{dong,chamel}, but have only stated it more circuitously. A more serious concern, however, is 
that they \cite{nityakon} have presented completely erroneous results, according to our understanding, 
which we correct in this Comment. 
Note that we use exactly the same symbols as in \cite{nityakon} for ease of comparison. Additionally, 
note that these authors \cite{nityakon} have used both $R_*$ and $R$ alternately to denote 
the stellar radius, however, in the present Comment we use only $R$ to avoid confusion.

We begin by stating that we have already solved the problem posed in \S IIA of \cite{nityakon} in 
our latest work \cite{jcap14}, where, by considering various magnetic field profiles, 
we have obtained stable magnetostatic equilibrium solutions of super-Chandrasekhar 
white dwarfs having mass as high as $3M_\odot$ in a general relativistic framework.

Next coming to \S IIB of \cite{nityakon}, we note that although the formulae used in 
this section are correct, 
\underline{all the numerical estimates 
are incorrect}! 
We point out that Table I of \cite{nityakon} lists incorrect values of 
${\cal Q} (R_m^g/R)^4$ corresponding to different $n$ as per our computation. However, before we correct their 
table and comment accordingly, first
we point out that even if one uses the {\it incorrect} values given in Table I of \cite{nityakon}, 
then also one does {\it not} arrive at the maximum allowed central field 
$B_{\rm upper-bound} \simeq 10^{16}$ G, as calculated in \cite{nityakon}, which we demonstrate below
to be highly underestimated. 
 
As we understand, the authors \cite{nityakon} have used Eqs. 17 and 21 of their work and the value of 
${\cal Q} (R_m^g/R)^4$ corresponding to $n=1$ given in Table I, to arrive at a value of $B_{\rm upper-bound}$ for 
the super-Chandrasekhar white dwarf in our work \cite{prl13}, having $M=2.58 M_{\odot}$, 
$R=69.5$ km, $B_c=8.8\times 10^{17}$ G and $R^g_m = 0.5 R$. 
Since, according to \cite{nityakon}, 
$P_{cm} = B_{\rm upper-bound}^2/24\pi$, one can write the expression 
for $B_{\rm upper-bound}$ using Eq. 17 of \cite{nityakon} as
\begin{equation}
B_{\rm upper-bound} \simeq \sqrt[]{24 \pi {\cal Q}(1) P_c} ~.
\end{equation}
Now, $P_c$ for the concerned super-Chandrasekhar white dwarf can be evaluated by using Eq. 19 
of \cite{nityakon}, noting that $\xi(R) = \pi$ and $\theta'(R)=-1/\pi$ for $n=1$ \cite{chandrastel}. 
Thus, one obtains $P_c=2.9\times 10^{32}$ erg/cc. One can also independently verify this value from 
Figure 1 of our work \cite{prl13}. Let us now consider 
${\cal Q}(1) = 2.1 (R/R_m^g)^4$, given in Table I of \cite{nityakon}, which the 
authors claim to have used in their calculation. Then, 
from equation (1) above, we obtain $B_{\rm upper-bound} \simeq 8.57\times 10^{17}$ G, which is 
practically the same as $B_c$. Thus, the apparently extensive effort put in 
by the authors seems to contradict their own conclusion!

Now coming back to the correction of Table I of \cite{nityakon}. We 
construct the correct table by using the solutions to the Lane-Emden equation for different $n$ 
and then putting them in Eq. 21 of \cite{nityakon}, which is easily verifiable.
As one can see, the values of ${\cal Q} (R_g^m/R)^4$ in Table I of \cite{nityakon} are exactly 
$10$ times higher than the correct values listed below!

\begin{table}[h]
\caption{}
\begin{center}
\begin{tabular}{|c|c|c|c|c|}
\hline
$n$ & $1$ &  $1.5$ & $2$ & $3$ \\
\hline

${\cal Q} (R_m^g/R)^4$ & $0.21$ & $0.189$ & $0.176$ & $0.162$ \\

\hline
\end{tabular}
\end{center}
\end{table}

However, even if we consider the 
{\it correct} value of ${\cal Q}(1) = 0.21 (R/R_m^g)^4$, obtained here, then also we arrive at
$B_{\rm upper-bound} \simeq 2.71\times 10^{17}$ G, 
which is still more than one order of magnitude higher than $10^{16}$ G and very close to 
the value of $B_c$. How the authors \cite{nityakon} have arrived at $B_{\rm upper-bound} \simeq 10^{16}$ G 
is a complete mystery! There only seems to be a mere factor of $3.2$ mismatch between $B_{\rm upper-bound}$ and $B_c$, 
which is similar to what stems from equipartition considerations, and this issue has already 
been addressed in our earlier work \cite{mpla14} (see \S2 and \S6 therein for details).
Note that the authors \cite{nityakon,nk2} have already referred to our above paper. However, if they would 
have carefully gone through our calculations, then they could have avoided the mistakes 
we point out in this Comment.

Moreover, the authors \cite{nityakon} have themselves admitted that the bound given by their Eq. 17 
can be weakened by invoking factors such as ``nonsphericity and a nonpolytropic 
running of pressure and density". 
Clearly, by taking the above factors into consideration, the correct value of $B_{\rm upper-bound}$ 
is likely to be even closer to $B_c$. 
Thus, the claim made by the authors that $B_{\rm upper-bound}$ obtained by them 
``is almost 2 orders of magnitude smaller than the field claimed to be present in
the center of such an object" simply falls apart!

Furthermore, in \S III of \cite{nityakon}, Eq. 23 is partly {\it incorrect}. It should be:
\begin{equation}
P_e = \frac{1}{2}n_e E_F = \frac{1}{2 \pi^2} \omega_c^2~.
\end{equation}
Note that, we have already addressed the issues regarding the role of magnetic density ($B^2/8\pi c^2$) 
and general relativistic effects, mentioned in \S III of \cite{nityakon}, 
in \S5 of our previous work \cite{mpla14} and more quantitatively in a recent work \cite{jcap14}.
Likewise, the concern regarding a possible neutronization is discussed in \S6 of our previous work 
\cite{mpla14} and in a recent work \cite{vishal}.

To summarize, we feel that Nityananda and Konar \cite{nityakon,nk2} have simply tried 
to put forth the same argument involving the virial theorem in a slightly fancier way, 
while contributing nothing new, in our view, to the discussion. Most surprisingly, 
they have arrived at an incorrect estimate of the maximum 
allowable central magnetic field for a super-Chandrasekhar white dwarf, which is supposedly 
the main (if not the only) result of their work!

We thank M.V. Vishal of IISc for cross-checking all the numerical calculations discussed here.
B.M. acknowledges partial support through research Grant No. ISRO/RES/2/367/10-11. 
U.D. thanks CSIR, India for financial support.

\end{document}